# Comparative Analysis of Single and Hybrid Neuro-Fuzzy-Based Models for an Industrial Heating Ventilation and Air Conditioning Control System

Sina Ardabili
*Institute of advanced studies Koszeg*
*University of Pannonia*
Koszeg, Hungary.
0000-0002-7744-7906

Bertalan Beszedes
*Alba Regia Technical Faculty Obuda University*
Budapest, Hungary.
beszedes.bertalan@amk.uni-obuda.hu

Laszlo Nadai
*Kalman Kando Faculty of Electrical Engineering Obuda University.*
Budapest, Hungary
0000-0001-8216-759X

Karoly Szell
*Alba Regia Technical Faculty Obuda University*
Budapest, Hungary.
0000-0001-7499-5643

Amir Mosavi [1,2*]
[1] *Department of Mathematics and Informatics, J. Selye University*
Komarno, Slovakia.
[2] *Bauhaus Universität Weimar*
Weimar, Germany
0000-0003-4842-0613

Felde Imre
*Kalman Kando Faculty of Electrical Engineering Obuda University*
Budapest, Hungary
0000-0003-4126-2480

*Abstract*—Hybridization of machine learning methods with soft computing techniques is an essential approach to improve the performance of the prediction models. Hybrid machine learning models, particularly, have gained popularity in the advancement of the high-performance control systems. Higher accuracy and better performance for prediction models of exergy destruction and energy consumption used in the control circuit of heating, ventilation, and air conditioning (HVAC) systems can be highly economical in the industrial scale to save energy. This research proposes two hybrid models of adaptive neuro-fuzzy inference system-particle swarm optimization (ANFIS-PSO), and adaptive neuro-fuzzy inference system-genetic algorithm (ANFIS-GA) for HVAC. The results are further compared with the single ANFIS model. The ANFIS-PSO model with the RMSE of 0.0065, MAE of 0.0028, and $R^2$ equal to 0.9999, with a minimum deviation of 0.0691 (KJ/s), outperforms the ANFIS-GA and single ANFIS models.

## I. Introduction

Machine learning has become essential for the advancement of novel control systems in various applications domains [1-3]. Machine learning methods are fast evolving to deliver more intelligent control systems with the higher performance [4-6]. Machine learning has been reported highly beneficial in the control systems of heating, ventilation, and air conditioning (HVAC) mechanisms [7-9]. Artificial neural networks (ANN), decision trees (DT), adaptive neuro-fuzzy inference system (ANFIS) and multilayer perceptron (MLP) are among the most popular machine learning methods used in HVAC control systems [10-15].

The recently proposed machine learning models for HVAC control systems are reported promising for energy saving and reducing energy deviation [16-20]. Thus, the improvement of machine learning models for higher accuracy and performance is essential [21-23]. However, the research is in the early stage, as the recent literature suggests a great potential in machine learning and many rooms to explore the application of new methods [24-27]. Soft computing techniques and optimization algorithms are shown beneficial in the preprocessing and postprocessing HVAC data [28-31]. However, the application of the hybrid machine learning models has been limited in this realm [11, 24]. Considering the higher performance reported in using hybrid machine learning models in other control systems, e.g., [32-35], a research gap is apparent in the advancement of HVAC control systems.

The contribution of this paper is to propose two new hybrid machine learning models, i.e., adaptive neuro-fuzzy inference system-particle swarm optimization (ANFIS-PSO), and adaptive neuro-fuzzy inference system-genetic algorithm (ANFIS-GA) to improve the performance of an HVAC control system. The results are to be compared with the previously proposed ANFIS model [15] to evaluate the performance of the hybrid model. Section two represents the description of the dataset and methods, and the results are presented in section three.

## II. MATERIALS AND METHODS

### A. Experimental data

The study's experimental data are gathered from an HVAC system used for the temperature-control mushrooms production room with a volume equal to 643.5 m³. Fig. 1 shows a schematic representation of the 5 parts of the system exergy adopted from [15]. Nine sensors, i.e., five platinum resistance thermometers, two board mount humidity detectors, and two manometers, are set to collect the temperature, relative humidity, and pressure data. The DAQMaster and data loggers are also used for sensor data management to configure parameters, real-time monitoring, and store the data. The detailed statement of the energy and exergy analysis of the HVAC system is available in [15].

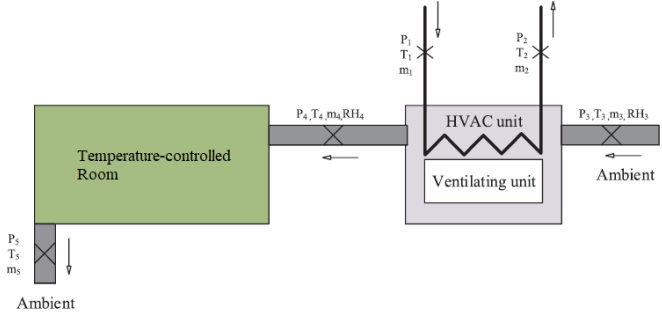

Fig. 1. Schematic representation of the HVAC unit used for the data collection

### B. Hybrid machine learning methods

The two hybrid methods of ANFIS-GA and ANFIS-PSO are used to develop the prediction models of exergy destruction and energy consumption. Both proposed methods have recently been gained popularity for advancing prediction models in a wide range of engineering applications including the control systems [36-39]. The ANFIS-GA hybridizes the components of a single ANFIS and genetic algorithm (GA) [40]. GA efficiently tunes the ANFIS controller through a global optimization represented by a set of ANFIS parameters, i.e., finding the optimum ANFIS parameters, see Fig. 2 [41]. ANFIS-GA has previously shown to outperform the accuracy of a single ANFIS model used in a control system [42]. Thus, it is expected that this hybrid form of ANFIS provides promising results.

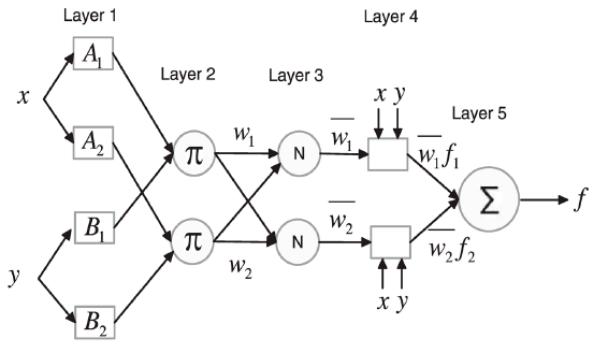

Fig. 2. Schematic representation of the ANFIS layers

In ANFIS, the model variables are constituted by the computation expressions of fuzzy logic and ANNs within the framework of five layers (see Fig.2). According to [43], in the first layer the $n$ nodes are defined based on the Gaussian functions as follows;

$$O_i^1 = \beta X = exp^{\left(-\frac{1}{2}\frac{x-z^2}{\sigma^2}\right)} \quad (1)$$

where O, Z, and $\sigma$, are output, the center of Gaussian function, and variance. The second and third layers ensure the accuracy of qualification and strength normalization of the model as follows;

$$O_i^2 = W_i = \beta_{Ai}X . \beta_{Bi}(X) \quad (2)$$

$$O_i^3 = \frac{W_i}{\sum W_i} \quad (3)$$

The fourth layer presents the effect of rules on outputs, where $n_i$, $m_i$ and $r_i$ represent the ANFIS linear parameters. The model further aims at reducing the difference between predicted values and experimental data as follows;

$$O_i^4 = \overline{w}_i f_{i=}\overline{w}_i(m_i X_1 + n_i X_2 + r_i) \quad (4)$$

In the last layer the weighted average summation is used to deliver a qualitative form of the model as follows;

$$O_i^5 = Y = \sum_i \overline{w}_i f_{i=}\overline{w}_1 f_1 + \overline{w}_2 f_2 = \frac{\sum W_i f_i}{\sum W_i} \quad (5)$$

Fig. 3 represents a schematic adaptation of the ANFIS-GA from [39], where a detailed description of the model flowchart is given.

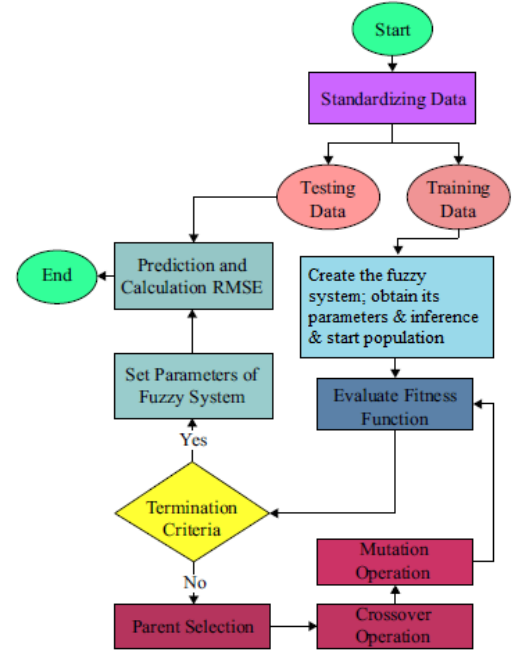

Fig. 3. Schematic representation of ANFIS-GA model

The second model is ANFIS-PSO to calculate the exergy destruction and energy consumption of the HVAC control system. ANFIS-PSO has recently shown outstanding results in enhancing the control systems [44]. Therefore, it has been chosen for this case study. ANFIS-PSO has been first proposed for the prediction of the wind power energy [45], where PSO represented a reliable approach for tuning the ANFIS parameters using a low number of variables for proper implementation, as presented in Fig. 4.

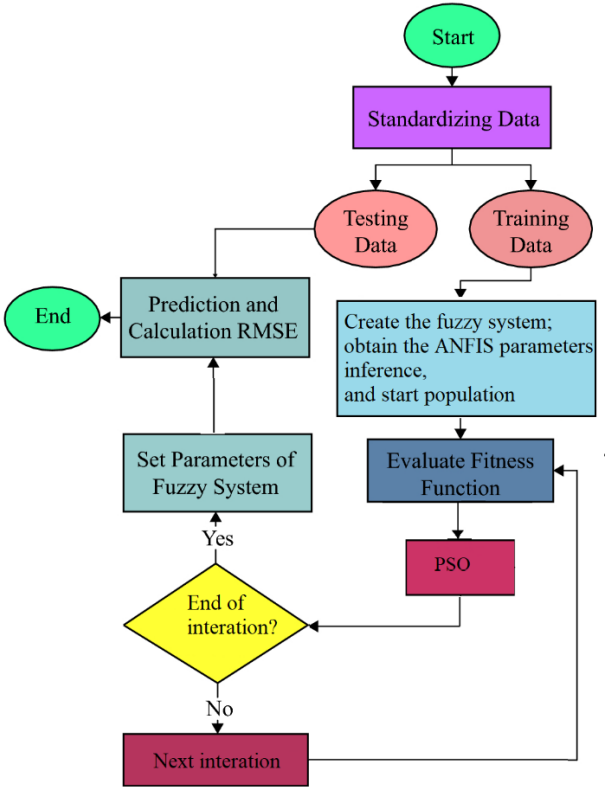

Fig. 4. Schematic representation of ANFIS-PSO model

The swarm in the PSO algorithm works based on random particles, which are the model solutions for tuning the ANFIS. Thus, the performance of each particle is assessed according to its fuzzy system. In ANFIS-PSO, the PSO is used to optimize the parameters of the fuzzy system as follows;

$R_i$: if $x_1(k)$ is $A_{i1}$ and … $x_n(k)$ is $A_{in}$, then $u(k)$ is $a_i$    (9)

where $x_n(k)$, $u(k)$, and $k$ represent the input variables, output variables, and time. And the $\vec{S_i}, \vec{V_i},$ and $\vec{P_i}$ represent the position, velocity, and vector of particles, respectively, where:

$$\vec{S_i}(t+1) = \vec{S_i}(t) + \vec{V_i}(t+1) \quad (10)$$

The optimum global value in the solution space of $P_s$ is $\vec{P_i^g}$ [46]. The PSO identifies the optimal antecedent parameters after the rule generation and initialization. The error-index $E(t)$ represents the evaluation function.

*C. Evaluation metrics*

The evaluation metrics of root mean square error (RMSE), Pearson correlation coefficient (R), and mean absolute error (MAE) are used to evaluate the performance of the models as follows;

$$RMSE = \sqrt{\frac{1}{N}\sum_{i=1}^{N}(A-P)^2} \quad (6)$$

$$R = \left(1 - \left(\frac{\sum_{i=1}^{n}(A-P)^2}{\sum_{i=1}^{n}A_i^2}\right)\right)^{1/2} \quad (7)$$

$$MAE = \frac{\sum_{i=1}^{n}|A-P|}{N} \quad (8)$$

In both stages of training and testing, the RMSE, $R^2$, and MAE values are calculated.

## III. RESULTS

The models are built based on the total exergy destruction of the system, i.e., output variable. The ambient temperature, air flow rate, and water flow rate ambient relative humidity are considered as independent variables, i.e., network inputs. The 70% of the data are used for training and 30% for testing. Table. I represents the training results for the three models. ANFIS-PSO shows better results compared to ANFIS and ANFIS-GA.

TABLE I. TRAINING RESULTS

| Method | Structure | RMSE | MAE | Deviation (KJ/s) |
|---|---|---|---|---|
| ANFIS | MF type: Gaussian Number of MFs: 3 Output: linear Optimizer type: hybrid | 0.024 | 0.014 | 0.3547 |
| ANFIS-GA | Max generation=282 Population size=200 | 0.005 | 0.0038 | 0.2123 |
| ANFIS-PSO | Max iteration=204 Swarm size=250 | **0.0017** | **0.00091** | **0.0502** |

Table. II represents the testing results for the three models. ANFIS-PSO shows better results compared to ANFIS and ANFIS-GA. Furthermore, Fig. 5 presents predicted values with $R^2$ for all the models. The comparative analysis of the deviation from the target value for the exergy destruction for all the models is given in Fig. 6 where ANFIS-PSO has delivered the minimum deviation.

TABLE II. TESTING RESULTS

| Method | Structure | RMSE | MAE | Deviation (KJ/s) |
|---|---|---|---|---|
| ANFIS | MF type: Gaussian Number of MFs: 3 Output: linear Optimizer type: hybrid | 0.068 | 0.04 | 0.9694 |
| ANFIS-GA | Max generation=282 Population size=200 | 0.0396 | 0.0226 | 0.5443 |
| NFIS-PSO | Max iteration=204 Swarm size=250 | **0.0065** | **0.0028** | **0.0691** |

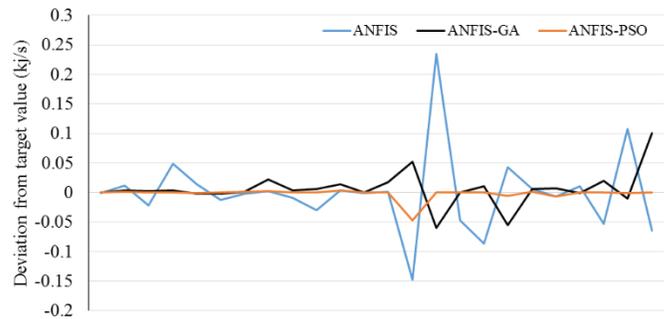

Fig. 6. Comparative deviation from the target value for the exergy destruction (KJ/s) for three models.

## IV. CONCLUSION

The hybridization of machine learning methods with soft computing techniques is an essential approach to improve the performance of the prediction models. Hybrid machine learning models, particularly, have gained popularity in the advancement of the high-performance control systems. Higher accuracy and better performance for prediction models of exergy destruction and energy consumption used in the control circuit of HVAC systems can be highly economical in the industrial scale to save energy. This research proposes two hybrid models of ANFIS-PSO and ANFIS-GA for the HVAC control system. The results are further compared with the single ANFIS model. The ANFIS-PSO model with the RMSE of 0.0065, MAE of 0.0028, and R2 equal to 0.9999, with a minimum deviation of 0.0691 (KJ/s), outperforms the ANFIS-GA and single ANFIS models. For the future research, advancement of hybrid and ensemble machine learning models, e.g., [47-52], and comparative analysis with deep learning models, e.g., [53-56] are proposed to identify models with higher efficiency.

ACKNOWLEDGMENT

We acknowledge the financial support of this work by the Hungarian State and the European Union under the EFOP-3.6.1-16-2016-00010 project and the 2017-1.3.1-VKE-2017-00025 project.

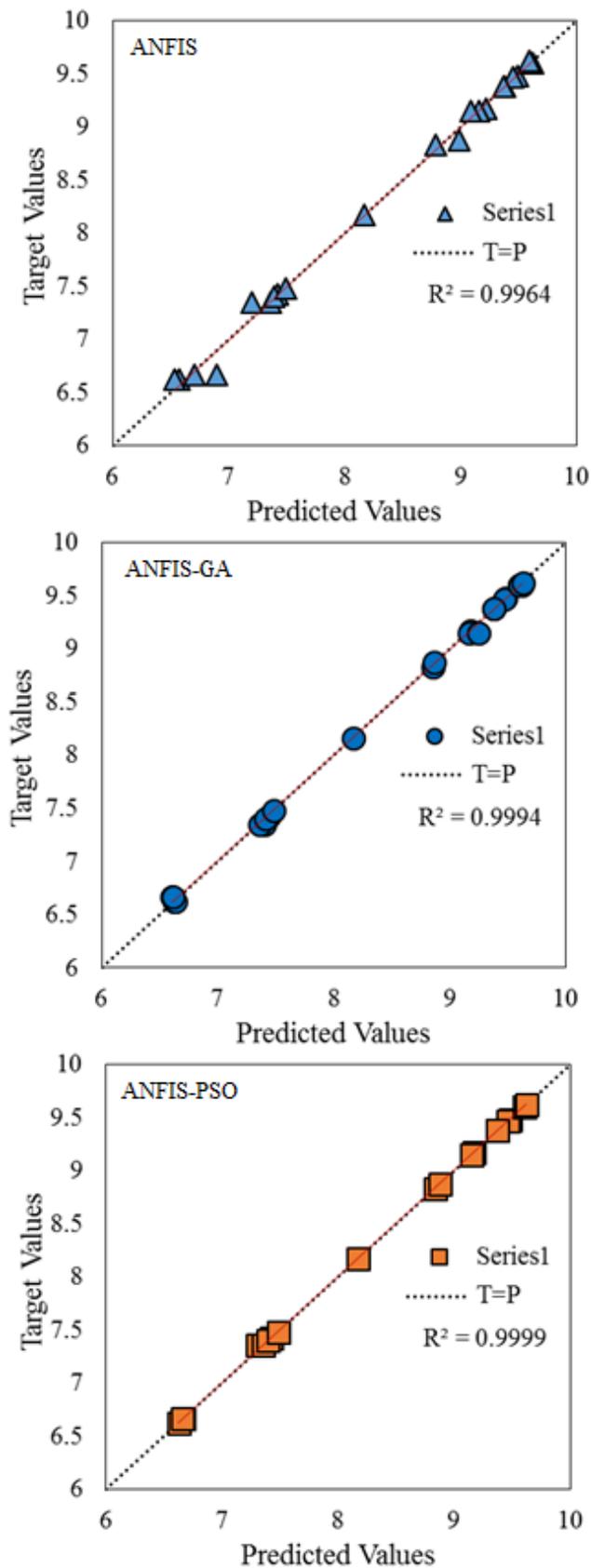

Fig. 5. Comparative analysis of the models for predicted values and $R^2$